# On the Relationship Between Proximal and Distal Stimuli and an Example of Its Significance to Physics

DOUGLAS M. SNYDER
LOS ANGELES, CALIFORNIA

ABSTRACT

The difference between proximal and distal stimuli is discussed and an example of the significance of this difference to physics is pointed out in the area of quantum teleportation. In particular, experimental evidence concerning the relative status of the distal stimulus to the observer points toward the relative nature of spatial orientation of objects. In quantum teleportation, this result indicates that the amount of classical information that needs to be transmitted to support the quantum transfer of information may be greater than is expected when the distal stimulus is assumed, as it ordinarily is, to have an absolute orientation in relation to the observer.

TEXT

In psychology, there is more than one kind of stimulus for the observer. Here we consider two kinds of stimuli involved in perception of the environment. The distal stimulus corresponds to what is generally considered the "actual" object in the environment. The proximal stimulus is generally defined as the pattern of energy impinging on the observer's sensory receptors. This energy is associated with a distal stimulus. The observer depends most directly on proximal stimuli, not distal stimuli, in perceiving his world.

For certain perceptions, there is little distinction between distal and proximal stimuli. Consider for example, touch. For this sense, the distal stimulus responsible for the pattern of stimulus energy is created when the object that comes to serve as the distal stimulus is in physical contact with the observer. With other senses, though, there may be a large distinction between the distal and proximal stimuli. For example, consider vision. In vision, the distal stimulus may be far from the observer and serve as the basis for stimulus energy that travels to the observer and that becomes the proximal stimulus when it reaches the retinas of the observer. We assume that there is a one-to-one correspondence between the proximal stimulus and distal stimulus, excepting for aberrations occurring while the stimulus energy is in route to the observer from the distal stimulus.

# On the Relationship

Experiments [1,2,3,4,5,6,7,8,9] have shown that this one-to-one correspondence need not exist, at least for vision. This work on vision has shown that more than one proximal stimulus can be associated with a single distal stimulus. These different proximal stimuli are ordinarily associated with distinct and unique distal stimuli.

That there may be more than one proximal stimulus associated with a single distal stimulus thus affects the nature of the distal stimulus. It cannot be held that the distal stimulus has a unique nature when we cannot in principle distinguish a unique proximal stimulus associated with it. It should be noted that whatever we know of distal stimuli is through proximal stimuli.

In general, it can be stated that if:

1) there are two different proximal stimuli corresponding to one distal stimulus,

2) no aberrations are responsible for there being two proximal stimuli, and

3) each proximal stimulus would ordinarily be associated with a distinct and unique distal stimulus,

there is no way to differentiate between which of the possible forms of the distal stimulus, each of would ordinarily be associated in a unique manner with the different proximal stimuli, is the "actual" one. Further, as noted, the distal stimulus is considered to be associated with the "actual" object in the world. One can surmise that the "actual" object in the world then has a possibility of at least two forms, the ordinary process of perceiving an object in the world associating a unique distal stimulus with each form. Actualizing one proximal stimulus, assuming a specific distal stimulus to be associated with it, and assuming a specific form of an object in the world that is the distal stimulus, involve cognitive actions on the part of the observer the results of which empirical evidence does not support. These results, though, serve to govern our interactions in the world, appearing to anchor them in a single ordered world instead of multiple worlds within which it seems it would be difficult to function.

Perception involves coordination of input from different senses, including vision. Visual perception, for example depends not only on sensory input to the eyes; it also depends on the coordination of input from other senses. The hypothesis that there may be more than one proximal stimulus associated with a distal stimulus is not affected by the coordination of input from different senses for perception in one sensory modality such as vision. For example, the experiments noted earlier were concerned with the orientation of the observer's visual field. Though orientation of the visual field is a feature of visual perception, this perceptual result depends on the coordination of information from other sensory modalities such as touch and balance in addition to information obtained through light impacting the eye.

The result that more than one proximal stimulus can be associated with a distal stimulus has important consequences for physics where the distal stimuli and the objects in the world with which they are associated are assumed to be independent of the observer. For example, consider quantum teleportation, a phenomenon that has received empirical support [10,11]. In quantum teleportation it is assumed that one might used fixed stars to determine the orientation of observers' spatial coordinate systems at points far distant from one another. It is assumed that the distal stimuli represented by the fixed stars are associated in a one-to-one manner with proximal stimuli representing the fixed stars at the retinas of the observers or at the measuring instruments used to gauge the orientation of the fixed stars, the readings of which can in turn be sent to the retinas of the observers via light. The measuring instruments themselves become a proxy for the orientation of the fixed stars with regard to the observer.

The assumption that one can rely on fixed stars for an absolute orientation of objects and coordinate systems in space with regard to an observer though does not hold when the experimental evidence from research on orientation of the observer's visual field is taken into account. In the experiments noted, where *all incoming light* is rotated by some constant degree or inverted along the horizontal, the observer demonstrates adaptation such that the visual field is perceived as upright. If there were an objective orientation of objects in space, how would an observer know it? Essentially, there appears to be a personal spatial structure used by an observer that in principle need not be uniform for all observers and which is tied to the orientation of objects in space for that specific observer [12]. In the case of quantum teleportation, more information might be required to align the coordinate systems of observers with one another than simply determining and communicating the orientation of the sender's coordinate system relative to the fixed stars to the observer receiving the teleported information. It is also possible that more than one way of aligning distant spatial coordinate systems relative to one another may be possible such that the same experimental results concerning teleportation may be obtained for what would ordinarily be considered different relative orientations of the distant coordinate systems from which the teleported information originates and in which the teleported information arrives.

### REFERENCES

[1]  G. M. Stratton. The Psychological Review, 3, 611 (1896).

[2]  G. M. Stratton. The Psychological Review, 4, 341 (1897).

[3]  G. M. Stratton. The Psychological Review, 4, 463 (1897).

[4]  T. Erismann., I. Kohler. Upright vision through inverting spectacles [Film]. University Park, Pennsylvania: PCR: Films and Video in the Behavioral Sciences (1953).

[5]  T. Erismann., I. Kohler. Living in a reversed world [Film]. University Park, Pennsylvania: PCR: Films and Video in the Behavioral Sciences (1958).

[6]  I. Kohler. Psychological Issues, 3, 19 and 165 (1964).

[7]  N. H. Pronko, N. H., F. W. Snyder. Vision with spatial inversion [Film]. University Park, Pennsylvania: PCR: Films and Video in the Behavioral Sciences (1951).

[8]  F. W. Snyder, N. H. Pronko. Vision with spatial inversion (University of Witchita Press, Witchita, Kansas, (1952).

[9]  H. Dolezal. Living in a world transformed. (Academic Press, New York, 1982).

[10] C. H. Bennett, G. Brassard, C. Crepeau, R. Jozsa, A. Peres, W. K. Wooters. Phys. Rev. Lett. 70, 1895 (1993).

[11] D. Bouwmeester, J. W. Pan, K. Mattle, M. Eibl, H. Weinfurter, A. Zeilinger. Nature 390, 575 (1997).

[12] D. M. Snyder. On the Positioning of Objects in Space. (1998). (xxx.lanl.gov.abs/physics/9909008, publish.aps.org/eprint/gateway/eplist/aps1998sep23_002).